# Label-free phase change detection of lipid bilayers using nanoscale diamond magnetometry

*Hitoshi Ishiwata\*, Hiroshi C. Watanabe, Shinya Hanashima,*
*Takayuki Iwasaki and Mutsuko Hatano*

Dr. H. Ishiwata. Author 1
PRESTO, Japan Science and Technology Agency, 7 Gobancho, Chiyoda,
152-0076 Tokyo, Japan
School of Engineering, Department of Electrical and Electronic Engineering, Tokyo Institute of Technology, 2-12-1 Ookayama, Meguro, Tokyo 152-8552, Japan
E-mail: ishiwata.h.aa@m.titech.ac.jp

Dr. H. C. Watanabe. Author 2
PRESTO, Japan Science and Technology Agency, 7 Gobancho, Chiyoda,
152-0076 Tokyo, Japan
Quantum Computing Center, Keio University, 3-14-1 Hiyoshi, Kohoku-ku,
Yokohama 223-8522, Japan

Prof. S. Hanashima. Author 3
Department of Chemistry, Graduate School of Science, Osaka University, 1-1 Machikaneyama, Toyonaka, Osaka 560-0043, Japan

Prof. T. Iwasaki. Author 4, Prof. M. Hatano. Author 5
School of Engineering, Department of Electrical and Electronic Engineering, Tokyo Institute of Technology, 2-12-1 Ookayama, Meguro, Tokyo 152-8552, Japan

Keywords: Nitrogen-vacancy center, nanoscale magnetometry, lipid bilayer

Abstract: The NV center in a diamond is a quantum sensor with exceptional quality for highly sensitive nanoscale analysis of NMR spectra and thermometry. In this study, we investigate nanoscale phase change detection of lipid bilayers utilizing ensemble-averaged nuclear spin detection from small volume ~ (6 nm)$^3$, which was determined by the depth of the NV center. Analysis of nanoscale NMR signal confirm thickness of lipid bilayer to be 6.2 nm ± 3.4 nm with proton density of 65 proton/nm$^3$ verifying formation of lipid bilayer on top of diamond sample. Correlation spectroscopy from nanoscale volume reveals quantum oscillation at 3.06 MHz corresponding to the Larmor frequency of proton at an applied magnetic field of 71.8 mT. The result of the correlation spectroscopy was compared with the 2D molecular diffusion model constructed by Monte Carlo simulation combined with results from molecular dynamics simulation. There is a change in diffusion constant from 1.5 ± 0.25 nm$^2$/μsec to 3.0 ± 0.5 nm$^2$/μsec when the temperature changes from 26.5 °C to 36.0 °C. Our results demonstrate that multi-parameter detection of changes in translational diffusion and temperature is possible in label-free measurements using nanoscale diamond magnetometry. Our method paves the way for label-free imaging of cell membranes for understanding its phase composition and dynamics.

**1. Introduction**

The cell membrane is a nanoscale 2D fluid crystalline assembly with sub-compartment domains that are critical for cellular functions, including transport of molecules, communications, and metabolic properties with its external medium.[1,2,3] These domains are distinguished by different phases of lipid membranes, and extensive research has focused on understanding the structure and dynamic properties of such domains.[4] The fluidity of the lipid bilayer, described by the 2D translational diffusion of lipid molecules,[5] determines the most fundamental property of lipids in different phases and therefore domains. Florescence microscopy has been most effective for measuring fluidity.[6, 7] Most advanced example includes Stimulated Emission Depletion-Fluorescence Correlation Spectroscopy (STED-FCS) utilized for detection of nanoscale diffusion and identification of nanoscale domain.[8] However, the use of fluorescent probes in such a technique changes the mass and structure of target molecules and deteriorates the observed dynamics.[9, 10] For direct measurement of the diffusion constant without additional perturbation in a biological environment, a label-free technique with nanoscale detection volume is necessary. Nanoscale NMR and correlation spectroscopy using a NV center has emerged as a quantum measurement platform that allows for label-free diffusion measurement with nuclear spin from a small detection volume of ~ (6 nm)$^3$.[11] The detection volume is determined by the location of the NV center from the surface of the diamond.[12, 13]

The NV Center also allows the detection of local temperature with sub-degree precision,[14,15] enabling the multi-parameter detection of temperature and diffusion in nanoscale samples in a biological environment, which is optimal for the phase change detection of biological samples. Our measurement technique is compatible with cell membrane measurement and can be achieved by simply placing cells on top of diamond. This measurement technique makes it possible to achieve imaging down to a diffraction-limited spot.

In this study, the NV center was formed as a perfectly aligned Delta doped layer[16] within 10 nm from the surface as a highly sensitive detection probe for the ensemble-averaged detection of nanoscale diffusion in the lipid bilayer. Correlation spectroscopy observed from the NV center revealed a change in relaxation time as a function of change in temperature. Monte Carlo 2D translational diffusion simulation was combined with molecular dynamics simulation to depict dynamics that are observed in our data. Our simulation demonstrated that the translational diffusion constant changed from $1.5 \pm 0.25$ nm$^2$/μsec to $3.0 \pm 0.5$ nm$^2$/μsec by changing the temperature from 26.5 °C to 36.0 °C. Our simulation demonstrates the phase change from an ordered phase to a rippled or disordered phase.

We report the first direct observation of phase change in the lipid bilayer; this was achieved using the nanoscale detection volume of the NV center. Direct observation of changes in the diffusion constant paves the way for the label-free identification of domains that are formed in the cell membrane to understand the relationship between cell membrane dynamics and cell function.[17]

## 2. Results
### 2.1. Experiment setup

Figure 1 (a) presents the overall setup of our experiment. A perfectly aligned shallow ensemble NV center was formed 10 nm from the surface through CVD growth.[16] NV center measurements were performed using a home-built optical microscope based on Olympus IX73 for confocal and wide-field measurements. Details are discussed in Experimental Section.

Supported lipid bilayer (SLB) was formed using dipalmitoylphosphatidylcholine (DPPC) molecules on top of the shallow NV center using the vesicle fusion method.[18] DPPC was used as the model in this study for the closeness of properties to sphingomyelin (SM), a major constituent of membrane rafts.[19] Phases in lipids are defined by fluidity of the lipid molecules at a given temperature, and therefore change in diffusion constant by change in temperature corresponds to change in phase of lipids.[20] As shown in Figure 1 (b), DPPC exhibits a solid ordered phase at ~25 °C and as the temperature is raised close to Tm (transition temperature) of DPPC ~41 °C, a phase change occurs in the DPPC toward the liquid disordered (Ld) phase.[21] To measure the change in diffusion rates of the lipid bilayer, an optically defined averaged readout from a shallow ensemble NV center with a detection volume ~ (6 nm)$^3$ was used for nanoscale NMR and correlation spectroscopy as shown in Supporting Information. Application of correlation spectroscopy using pulse sequence shown in Figure 1 (a), allows comparison of the detected phase accumulated in the NV center between two XY8-N measurements spaced by tau τ. Correlation spectroscopy has been shown to detect 3D nanoscale diffusion of protons in oil.[11] In this study, we applied this technique to high density (~60-70 nm$^{-3}$) proton nuclear spins in a lipid bilayer to study the diffusion characteristics, as shown in Figure 1 (b).

The sample was placed in an incubator as shown in Figure 1 (a) to control temperature and maintain a steady temperature. For precise interpretation of phase transition, high-precision measurement of temperature is extremely important. Detection of the splitting between spin state $|0\rangle$ and $|\pm 1\rangle$ states enables readout of the resonance frequency D(T) that depends on the local temperature T. The temperature dependence is dD(T)/dT = -74 kHz/K.[22] We applied the pulse sequence shown in Figure 1 (a) known as Thermal Echo (T-Echo)[23, 24] to determine D(T) of the NV center by changing the applied frequency and measuring the observed oscillation frequency.

**2.2. Measurement confirmation of the lipid bilayer on the diamond surface.**

The formation of the lipid bilayer on the diamond sample was confirmed through florescent recovery after photobleaching (FRAP) performed with DPPC sample mixed with 1% mol Rhodamine B. The sample was excited with a 532 nm laser with a dichroic mirror (LPF at 600 nm) using a 40x objective lens (Olympus LCACHN40XIPC) and detected with a color camera (V230CFL or DP53). Figure 2 (a) depicts a typical image of a diamond sample with excitation with a 532 nm laser. Emission above 600 nm from Rhodamine B was confirmed on top of the diamond. Laser excitation was applied at a high intensity of ~$10^5$ mW/cm$^2$ for more than 5 min to perform photobleaching, where emission from Rhodamine B was depleted. Figure 2 (b) presents a typical image obtained after bleaching of the sample. Depletion of emission was confirmed on the spot defined by Iris used to block the laser. As shown in Figure (c), the DPPC/Rhodamine-PE sample showed recovery of emission after 5 min. FRAP measurement demonstrates recovery of emission on laser damaged spot of lipid bilayer, which confirms two things; Rhodamine B is bleached in the area of high laser exposure and lipid bilayer stayed intact as a bilayer on top of diamond substrate. Continuity of lipid bilayer provides fluidity that allows recovery of florescence over 5 min. In the case of randomly deposited lipids, the laser damaged spot shows no recovery because no diffusion is observed between damaged spot and non-damaged spot due to absence of continuous lipid bilayer. FRAP measurement proved the existence of SLB composed of DPPC molecules on top of the diamond sample.

Nanoscale NMR measurements were performed using a confocal setup for protons in the DPPC sample without rhodamine B to confirm the existence of DPPC on top of the shallow NV center. Sample preparation was applied in the same manner as DPPC/Rhodamine-PE described above, except that Rhodamine-PE was not introduced, and only DPPC (1,2-Dipalmitoyl-3-PC,13C32, Larodan) was used as the molecule for deposition. DI water was used instead of PBS solution to avoid uncertainty caused by interaction between PBS contents and microwave excitation that is used for NV measurement. DI water is also commonly used for preparation of lipid bilayer and no effect on change in phase transition for change in pH or sodium chloride concentration has been reported in previous study.[25] For nanoscale NMR measurements, a magnetic field of 71.8 mT was applied and confirmed from the ODMR spectrum of the NV center. The proton signal was observed by application of the XY8-40 sequence. A schematic representation of the estimation of lipid bilayer thickness $t_{LB}$ is presented in Figure 2 (d). The proton density and proton thickness of the lipid bilayer were calculated using the equation below with α=0 for NV centers oriented toward [111] direction as stated in Experimental Section.[12] $C(\tau)$ in equation (1) represents normalized contrast observed in nanoscale NMR measurement.

$$C(\tau) \approx \exp[-\frac{2}{\pi^2}\gamma_e^2 B_{RMS}^2 K(N\tau)] \, , \, B_{RMS\,calib.}^2 = \rho \, (\frac{\mu_0 \hbar \gamma_n}{4\pi})^2 (\frac{\pi[8-3\sin^4\alpha]}{128 d_{NV}^3}) \quad (1)$$

$$B_{RMS\,lipid\,bilayer}^2 = \rho \, (\frac{\mu_0 \hbar \gamma_n}{4\pi})^2 (\frac{\pi[8-3\sin^4\alpha]}{128})[\frac{1}{(d_{NV})^3} - \frac{1}{(d_{NV}+t_{LB})^3}] \quad (2)$$

ℏ Initially, the depth of the NV center was calibrated by measuring the protons in oil and using Eq.1 to evaluate the depth of the NV center to be 6.6 ±0.5 nm. Proton density inside lipid bilayer has been calculated by fitting data obtained from Radially Distributed Function to be 65 [proton/nm$^3$]. Details for calculation of uniform density is shown in Supporting Information Fig.2S. Using the calibrated and calculated value, the thickness of the lipid bilayer and the density of protons in the lipid bilayer were calculated using Eq.2. The calculated thickness of the lipid bilayer was estimated to be 6.2 ±3.4 nm with a proton density of 65 [proton/nm$^3$]. The lipid packing density estimated from proton density ~0.49 [nm$^2$/lipid] is comparable to the reported lipid packing density 0.47 [nm$^2$/lipid] for DPPC.[26] Our measurement demonstrates observation of proton density and thickness that are comparable to reported values from DPPC on top of the diamond sample. Details for calculation of lipid packing density is shown in Supporting Information "Calculation of proton density in lipid bilayer and conversion to lipid packing density" section. Measurement of nanoscale NMR was performed using confocal microscope with diffraction limited spot size of ~300nm. And measurement was performed on terrace section of diamond with flatness of less than ~2nm.

**2.3. Modeling of lipid bilayer and observation of phase change.**

Calculation of diffusion constant from correlation spectroscopy requires detailed modeling of dynamics for observed nuclear spin. A previous study demonstrated a 3D diffusion model for proton nuclear spin in oil.[11] The relaxation rate observed in correlation spectroscopy is determined by two factors, T$_2$ of observed nuclear spin and probability of detecting nuclear spin within the detection volume determined by the depth of the NV center. Because we observe proton spins with the S=1/2 system, the contribution to T$_2$ is determined by dipolar coupling between protons.[27, 28] As shown in Figure 3(a), the molecular dynamics of the lipid bilayer were divided into intramolecular and intermolecular parts, where the intramolecular part was used to model the rotation and wobble effect, and the intermolecular part was used to model the effect of diffusion. Equations used for calculation of T$_2$ are given below[28]: where r is the intramolecular distance between two protons in the lipid molecule, $\gamma_H$ is the proton gyromagnetic ratio, $\omega_H$ is the proton Larmor frequency, ℏ is the Dirac constant, $\tau_{c\,rot.}$ is the correlation time for rotation, N is the density of proton in the lipid bilayer, d is the intermolecular distance between two protons, and $\tau_{c\,trans.}$ is the correlation time for translation.

$$\frac{1}{T_{2\,intra.}} = \frac{3}{20}\frac{\gamma_H^4 \hbar^2}{r^6}\left[3\tau_{c\,rot.} + \frac{5\tau_{c\,rot.}}{1+\omega_H^2 \tau_{c\,rot.}^2} + \frac{2\tau_{c\,rot.}}{1+4\omega_H^2 \tau_{c\,rot.}^2}\right] \qquad (3)$$

$$\frac{1}{T_{2\,inter.}} = \frac{2\pi}{10}\frac{N\gamma_H^4 \hbar^2}{d^3}\left[3\tau_{c\,trans.} + \frac{5\tau_{c\,trans.}}{1+\omega_H^2 \tau_{c\,trans.}^2} + \frac{2\tau_{c\,trans.}}{1+4\omega_H^2 \tau_{c\,trans.}^2}\right] \qquad (4)$$

As shown in Figure 3(b), molecular dynamics (MD) simulation was used for dynamics in a small time scale of ~50 ns to calculate r, d, N, and $\tau_{c\,rot.}$. r, d, and N were all estimated from the integrated radially distributed function (RDF) (Supporting Information Fig.S2 (b),(c)). $\tau_{c\,rot.}$ was determined by estimating the correlation time for vector defined along the acid chain of DPPC (Supporting Information Fig.S2(d)). The results from MD simulation were combined with Monte Carlo simulation to calculate the change in the probability of detection in nanoscale magnetometry with a diffusion constant as the only free parameter for fitting to our model. The change in the probability of detection was calculated using a 2D diffusion model, where the overlap between the diffusion area and detection area was calculated to extract the change in the probability of diffusion (Supporting Information Fig.S3(c) and (d)). The detection area was calculated in detail using a model from a previous report[13] (Supporting Information Fig.S1(a) and (b)). In the Monte Carlo simulation, ~1nm proton layer observed on acid cleaned diamond surface[11, 16, 29] has been included as an immobile proton layer on the bottom of lipid bilayer[11]. This assumption is consistent with neutron scattering results on existence of proton layer on the bottom of lipid bilayer[30, 31].

For measurement of the phase transition in DPPC molecules, the temperature setting in the incubation chamber was used to change the temperature, and the NV center was used as a local probe for quantum thermometry using the Thermal-Echo (T-Echo) method. Pulse sequence was applied in a confocal setup at dual frequency to remove the effect of the magnetic field during measurement for precise measurement of $D(T)$ values below 1 °C precision. The results are shown in Figure 3 (c) and (d), where the energy difference between the bright |0> and dark |±1> states were observed as an oscillations in the T-Echo signal. The T-echo signal without heat supply from the incubator resulted in a $D(T)$ value of +0.10 MHz ± 24 kHz with respect to theoretical value of 2.87GHz. The temperature at this measurement setting was calibrated with a K class thermocouple (WT 100) to be 26.5 °C. T-Echo measurement with a temperature setting of 45 °C in an incubator resulted in a $D(T)$ value of -0.59 MHz ± 44 kHz with respect to theoretical value of 2.87GHz. The difference in the value of $D(T)$ between the two temperature settings provides a temperature difference of 9.45 °C with a 0.47 °C temperature precision determined by the distribution of the value for different frequencies of the applied pulse on T-Echo. All measurements at different temperature were carried after at least 12 hours of waiting time to stabilize temperature.

For both 26.5 °C and 36.0 °C, correlation spectroscopy was performed using an EMCCD camera as the detector for diffusion analysis of protons in DPPC molecules. The ODMR spectrum obtained from the NV center was used to determine the applied magnetic field of 71.8 mT. Example of result obtained from correlation spectrum for 1.6 μsec to 20 μsec is shown in Figure 3 (e). As shown in Figure 3 (e) oscillation at 3.06 MHz corresponding to Larmor frequency of proton was observed at 26.5 °C and 36.0 °C. The obtained data are compared with Monte Carlo simulation results in Figure 3 (e). As shown in Figure 3 (e), the correlation spectrum obtained at 26.5 °C relaxation characteristic shows a comparable character to the simulation with a diffusion constant Dt of 1.5 nm$^2$/μsec. For detailed comparison of obtained data and Monte Carlo simulation, absolute value of correlation spectrum at 26.5 °C and 36.0 °C are plotted in Figure 3 (f) and compared with results of 2D molecular diffusion simulation. Figure 3 (f) shows as the temperature is increased to 36.0 °C, the relaxation characteristic of correlation spectrum shows result similar to the simulation with a diffusion constant of 3.0 nm$^2$/μsec. And correlation spectrum obtained at 26.5°C shows relaxation characteristic similar to the simulation with a diffusion constant of 1.5 nm$^2$/μsec. Difference in observed diffusion constant at different temperature is clearly demonstrated by comparison with simulation. And relaxation characteristics of simulation with different diffusion constant are clearly distinguished. Accuracy of this measurement was tested by measuring diffusion constant of DPPC molecules when temperature of sample was once again lowered to 26.5 °C. The result is shown in Figure 3 (f) as 26.5 °C 2$^{nd}$ which shows complete overlap on relaxation characteristic with 26.5 °C 1$^{st}$ measurement demonstrating reversibility of phase transition with measured diffusion constant of 1.5 nm$^2$/μsec. This result is consistent with previous reports on measurement of reversibility of phase transition in lipid bilayer[32]. From correlation spectroscopy and comparison with 2D Monte Carlo simulation, diffusion constant of DPPC molecules have been estimated to be $1.5 \pm 0.25$ nm$^2$/μsec at 26.5 °C and $3.0 \pm 0.5$ nm$^2$/μsec at 36.0 °C with an error bar included for uncertainties (e.g. exact thickness of proton layer and exact location) caused by proton layer on the bottom of lipid bilayer. All measurements were performed on same location, calibrated at diffraction limited resolution (~300nm) by confocal scan of step-edge structure[16] observed on shallow ensemble NV centers. These measurements confirmed observation of phase change through change in the diffusion constant of DPPC molecules.

## 3. Discussion

Figure 4 presents the diffusion constants obtained from different techniques on DPPC molecules.[33, 34, 35, 36] Techniques that involve a florescent marker, such as FRAP and Marker Quench, exhibit diffusion constants that have an order of magnitude lower than that of label free techniques, such as magic angle spinning pulse field gradient NMR (MAS PFG NMR). In previous studies, use of florescent marker raised question on influence of structural changes induced by the marker.[37, 38] In addition, it has been reported that results from FRAP vary significantly depending on the selected markers,[38] and this is demonstrated by comparing two experiments performed on SLB of DPPC, as shown in Figure 4. The diffusion constant obtained from L. K. Tamm et al. with the NBD-PE marker differed significantly from the results of C. Scomparin et al. with the 16:0-12:0 NBD PC marker.

Our results obtained from nanoscale diamond magnetometry are consistent with the results obtained from MAS PFG NMR. MAS PFG NMR allows precise diffusion measurements without special labeling of the investigated molecules, which may alter molecular properties.[39] However, uniformity on structure of lipid membranes are necessary for modeling and radius curvature of membranes have to be larger than 1 μm due to the limitations imposed by the field gradient applied for measurement. Therefore, MAS PFG NMR does not allow direct measurement of actual cell membranes. Our measurement technique, using extremely small detection volume defined by the depth of the NV center, allows label-free measurement of diffusion constants from cell membranes without complication imposed by sample preparation and measurement system. This technique paves way for diagnostics of cell membranes, where imaging of fluidity on each individual cells are necessary.

## 4. Conclusion

In this paper, we report determination of diffusion constant of a lipid bilayer, a biological parameter that determines the dynamics of the lipid bilayer, by making use of extremely small detection volume offered by nanoscale NMR. Observation of diffusion constant reveals different phases of lipid bilayer which identifies sub-compartment domains that are critical for cellular functions. Our method builds foundation for label-free imaging of cell membranes for observation of phase composition and pristine dynamics that determines cellular functions.[17]

## 5. Experimental Section

*NV measurement system*:

The breadboard was placed inside Olympus IX-73 to guide the laser into the objective lens through a dichroic mirror. The detector side could be switched to an EM-CCD camera (iXonUltra) or a pinhole with an APD detector (SPCM-AQRH-14-FC-ND) or a color CCD camera (V230CFL or DP53) through the adjuster placed on the lower deck of IX-73. The incubator was placed inside the piezo stage (P-545.3C8S) to control the temperature with a thermocoupled-heater in order to change the temperature of our system. MW was delivered to the NV center through a 20 um diameter copper wire with a sputtered Ti/Cu/Au electrode on a cover slide. Microwaves were delivered from the SG (Anritsu MG3700A and SynthHD PRO Dual RF Signal Generator) combined with an amplifier (R&K CGA701M62-444R or Amplifier Research Model 50W1000A). The pulse sequence was controlled by DTG5274 and the MW pulse was truncated by a switch (Mini circuit ZASWA-2-50DRA+). A high-power laser (Verdi

G5) was pulsed through an AOM(Gooch Housego Model:3250-220) with an RF driver (3910-XX).

*NV center sample (definition of ensemble average readout)*:

Shallow NV ensemble was formed within 10 nm from the surface by the CVD growth technique on [111]-oriented diamond with step flow growth. [16] NV centers that are used in this experiment are oriented toward [111] direction. Diamond surface could be categorized into two parts, step edge structure and terrace section. [16] Overall surface roughness of 2-6nm induced by step edge structures and surface roughness of less than ~2nm is achieved on terrace section of sample surface.

In the case of shallow ensemble NV center, performing nanoscale NMR or correlation spectroscopy results in each individual single NV center detecting detection volume at depth defined by depth of NV center. In the optical readout process, each phase detected at individually independent (ensemble) NV centers are readout as averaged from optically defined detection area. This type of detection is generally regarded as an ensemble average detection of NV center which has been generally accepted to accommodate with single NV center model in the previous study[29]. And in the previous report[16], we demonstrated that results are comparable when same exact sample that is measured with single NV center in reference 29 of the manuscript is measured with perfectly aligned shallow ensemble NV center used in this manuscript.

*Laser spot size and number of NV center used for each measurement*:

In this paper, optically defined area was ~300nm diameter defined by optical diffraction-limited FWHM of Gaussian laser profile. Density of NV centers used in this experiment are ~ $10^{16}$ cm$^{-3}$ corresponding to ~ 10 NV centers per optically defined detection spot for the case of confocal microscope.

Spot size of laser used for EMCCD camera detection was 120um diameter and florescence was projected to EMCCD camera with ~300nm spatial resolution on each pixel.

EMCCD camera was used for all correlation spectroscopy for better signal to noise ratio observed by obtaining data averaged over 20 pixels by 20 pixels on EMCCD camera. Number of NV center used to detect data for correlation spectrum using EMCCD camera is ~4000 NV centers.

*Preparation method of DPPC supported lipid bilayer*:

Lissamine Rhodamine B 1,2-Dihexadecanoyl-sn-Glycero-3-Phosphoethanolamine (Rhodamine-PE, Thermo Fisher Scientific) was mixed at 1 % mol with 1,2-dipalmitoyl-sn-glycero-3-phosphocholine (DPPC, Avanti Polar Lipids) in PBS buffer. DPPC was dissolved in PBS at 0.5 mg/mL concentration and sonicated in a warm water bath at 60 °C for 5–15 min. After sonication, the sample was incubated for at least 1 h on a hot plate at 70 °C. After incubation, the sample was sonicated in a warm water bath at 60 °C until a transparent sample was obtained. CVD-grown diamond samples with a shallow NV center were exposed to an acid treatment with a 1 : 3 mixture of $HNO_3$ and $H_2SO_4$ at a hotplate temperature of 400°C for 45 min, followed by rinsing with deionized water to obtain an oxygen-terminated surface. XPS measurements of the acid-treated diamond samples confirmed increased oxygen coverage for confirmation of the oxygen termination of the surface. 100 μL of DPPC/Rhodamine-PE solution was deposited on top of diamond sample and incubated for 20 min. After incubation, 1 mL of PBS solution was injected onto diamond, and sample wash was performed ten times.

*MD simulation*:

The MD simulations were conducted using the Fujitsu PRIMEGY CX600M1/CX1640M1 (Oakforest-PACS) and SGI Rackable C1102-GP8 (Reedbush).

**Supporting Information**
Supporting Information is available from the Wiley Online Library or from the author.

**Acknowledgements**
This research was supported by JST PRESTO Grant number JPMJPR17G1 and JPMJPR17GC, JSPS Grant Number 20K03885 and 20H05518 and the MEXT Quantum Leap

Flagship Program (MEXT Q-LEAP) Grant Number JPMXS0118067285 and JPMXS0118067395.

H.I. conceived the concept, built the experimental setup for nanoscale NMR, correlation spectroscopy, and T-echo, fabricated shallow ensemble NV center, fabricated lipid bilayers, performed the measurements, built Python code for Monte Carlo 2D molecular diffusion analysis, analyzed the data, and wrote the manuscript. H.W. built and performed the MD simulation. S.H. advised lipid bilayer discussion and preparation. H.I., H.W., S.H., T.I., and M.H. discussed the study.

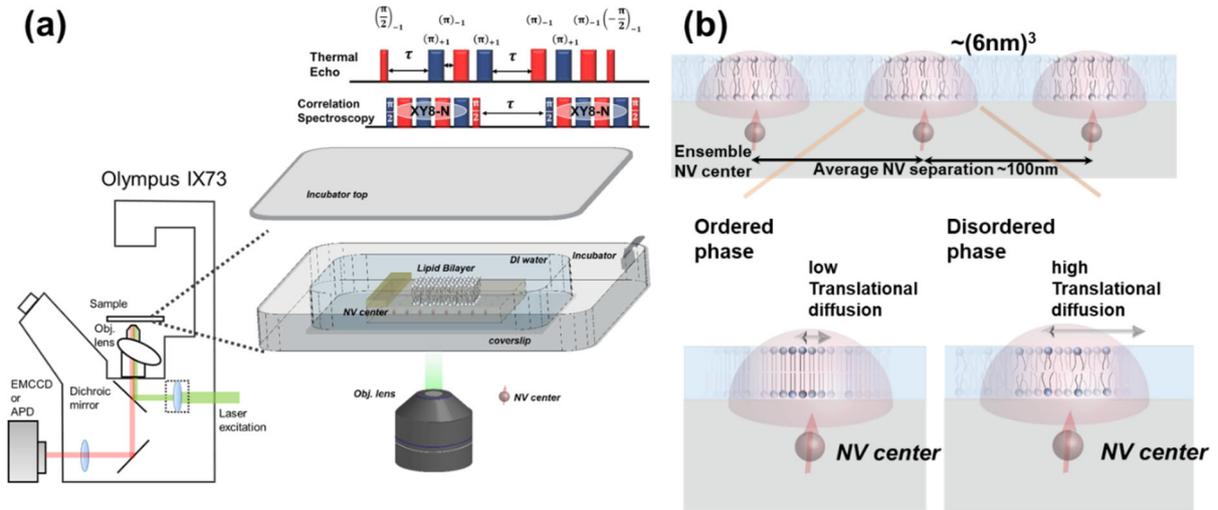

**Figure 1.** (a) Schematic for NV measurement system and used pulse sequences. For Thermal Echo blue and red corresponds to pulse that is used to address +1 and -1 states of NV center. For Correlation Spectroscopy red and blue corresponds to X and Y phases of pulses that are used in sequence. (b) Schematic for phase change detection of lipid bilayer using shallow ensemble NV center.

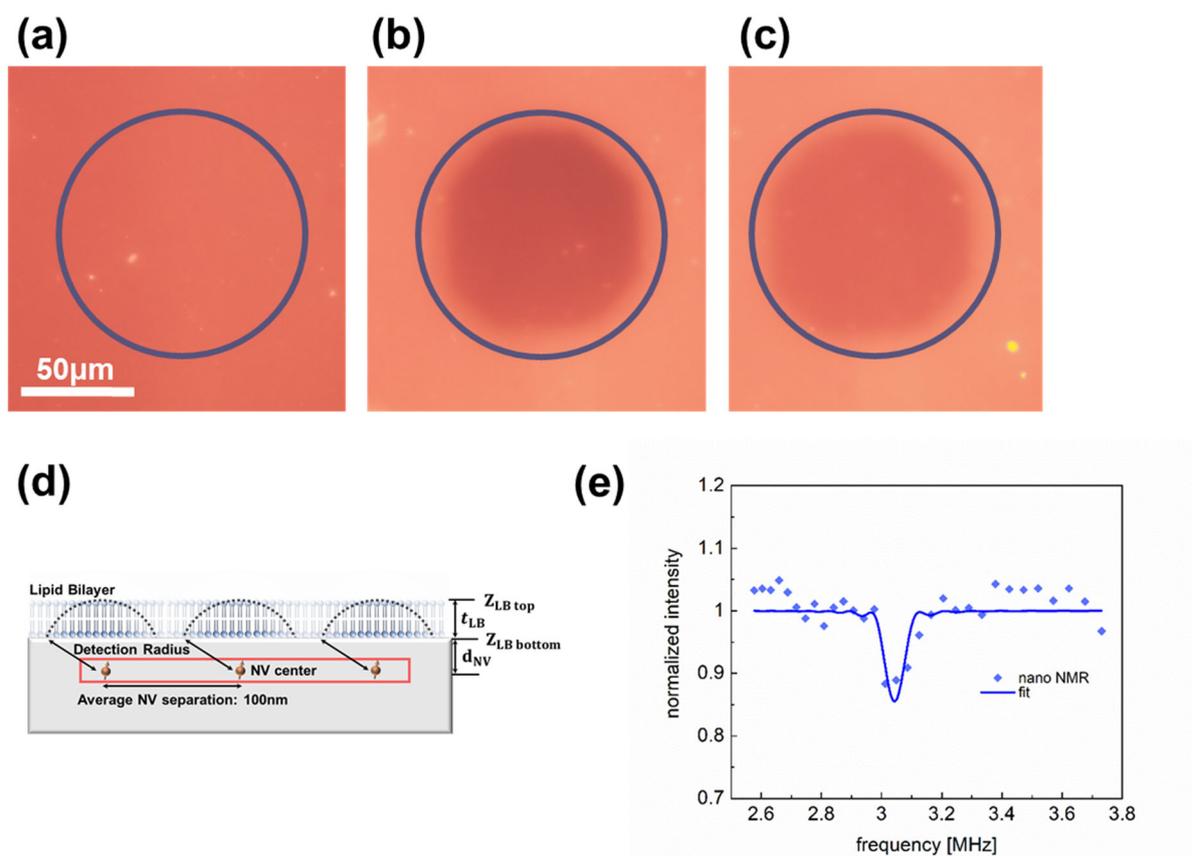

**Figure 2.** Image obtained with color camera on DPPC/Rhodamine-PE (a) after deposition (b) after photo bleaching (c) 5min. after photo bleaching. Blue circle in each picture represents laser spot that is applied. (d) Schematic representation for definition used to calculate thickness and density of Lipid Bilayer (e) Nanoscale NMR measurement performed on DPPC with applied magnetic field of 71.8 mT.

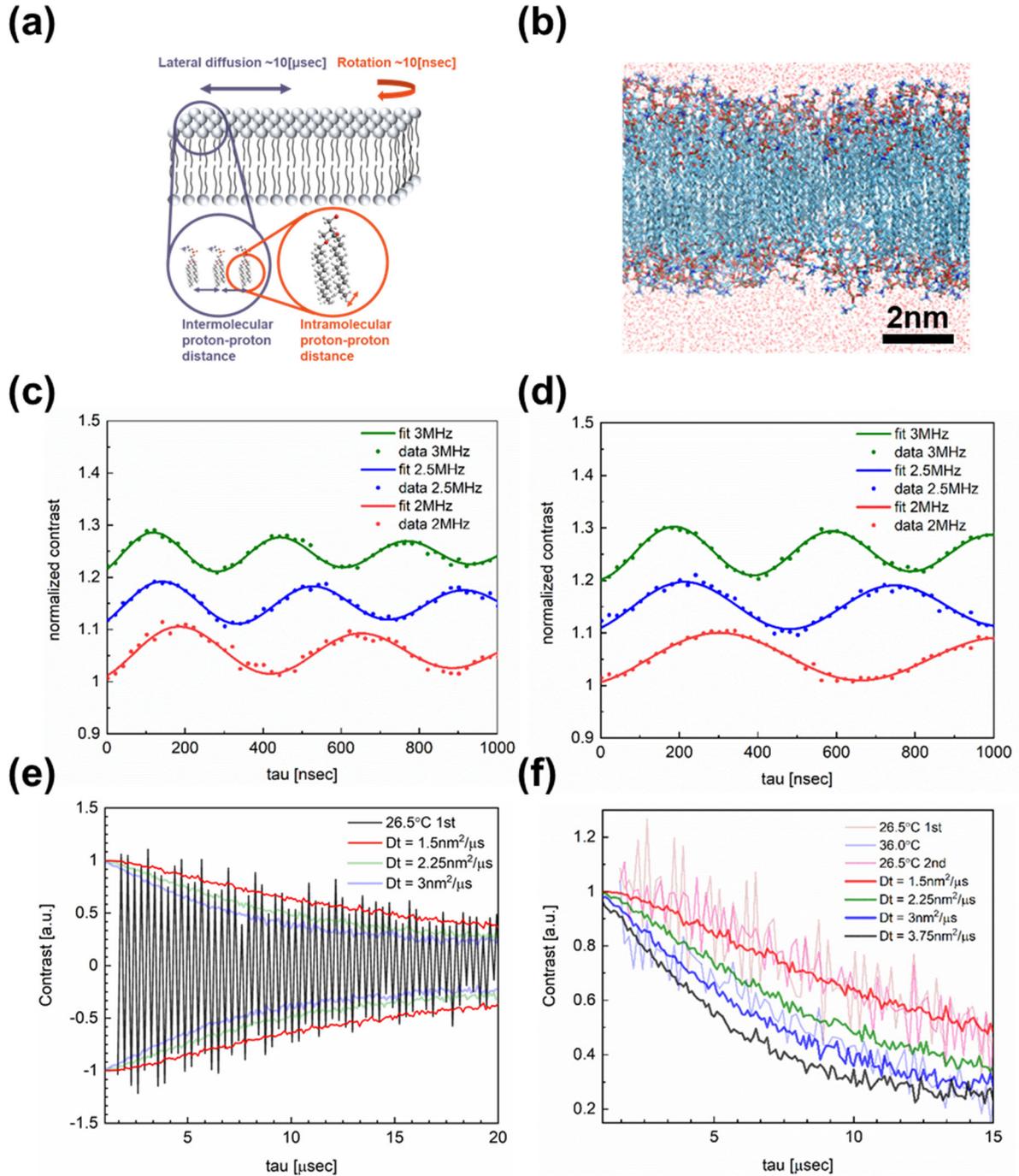

**Figure 3.** (a) Schematic for modeling of lipid bilayer. Lateral diffusion and rotation were separated for different time scales. Dipolar interaction was conceived for each dynamic by considering intermolecular and intramolecular proton-proton distance. (b) Molecular dynamics simulation on DPPC at 25 °C. Simulation was performed for 50 ns with 1 psec resolution. (c) T-Echo measurement performed at 26.5 °C for different set frequency relative to D(T) value. (d) T-Echo measurement performed at 36.0 °C for different set frequency relative to D(T) value. Correlation spectroscopy performed at (e) 26.5 °C and (f) 36.0 °C.

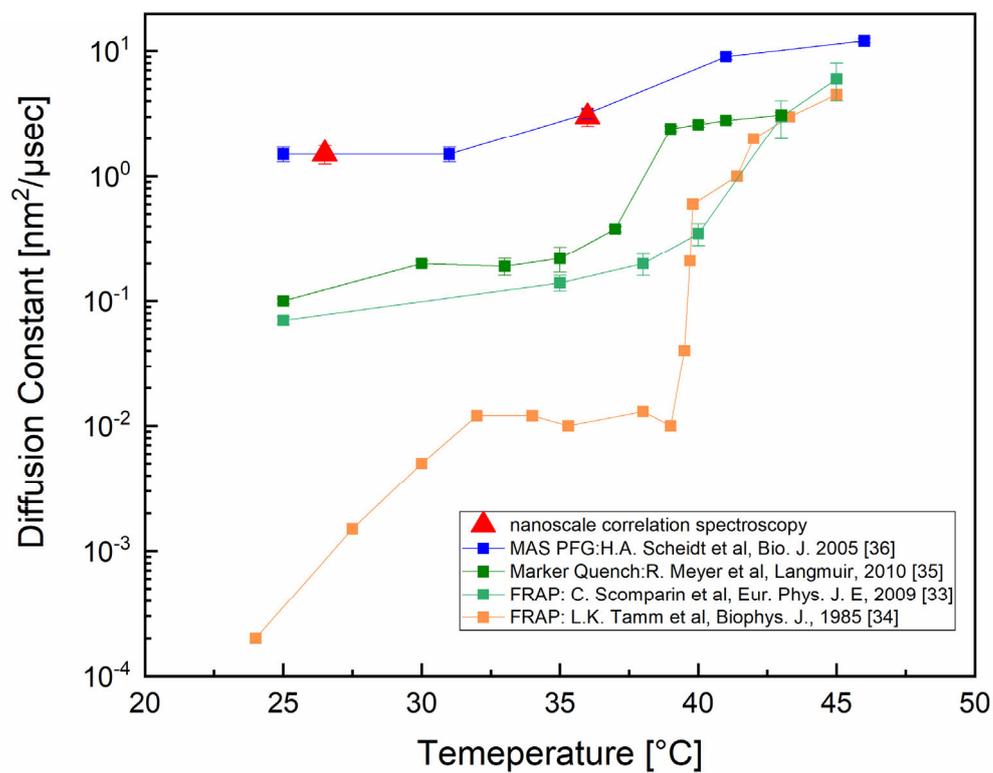

**Figure 4.** Comparison of diffusion constants obtained for different temperatures with different measurement techniques on DPPC molecules. Error bars were provided as indicated in each manuscript.

# Supporting Information

**Label-free phase change detection of lipid bilayers using nanoscale diamond magnetometry**

*Hitoshi Ishiwata\*, Hiroshi C. Watanabe, Shinya Hanashima, Takayuki Iwasaki and Mutsuko Hatano*

**Definition of detection volume of NV center:**

The detection volume for the shallow ensemble NV center was calculated following Monte Carlo simulation used to calculate the detection volume used in previous work (ref. 13). The density of proton in the lipid bilayer was estimated from molecular dynamics simulation and results obtained from nanoscale NMR measurements using NV center. Monte Carlo simulation was performed under the assumption of ~5nm thickness lipid bilayer and no detectable proton above lipid bilayer. (proton in DI water is ignored because of its extremely high diffusion coefficient). The following equations were used to calculate the detection volume for the case of nanoscale proton detection in the lipid bilayer, as shown in Figure S1 (a) and (b). The detection volume is defined as the volume of proton contributing to 70% of the total signal detected at the depth of the NV center.

$$B_{rms,i} = \sqrt{\tfrac{2}{3}}\left(\tfrac{3\mu_0\mu_p}{4\pi r_i^3}\right)\left(\tfrac{5\pi}{96}\right)\sqrt{(\vec{e_2}\cdot\vec{n_l})(\vec{e_1}\cdot\vec{n_l}))^2 + (\vec{e_3}\cdot\vec{n_l})(\vec{e_1}\cdot\vec{n_l}))^2}\,] \qquad [S1]$$

$$B(N) = \sqrt{\sum_{i=0}^{N} B_{rms,i}^2} \qquad [S2]$$

The number of protons contributing to 70% of the total signal was estimated as shown in Figure 1 (b), and the calculated number of protons was converted to the detection volume and detection area by assuming a thickness of ~ 5 nm for the lipid bilayer. The detection area was found to be 6.3 nm in radius from these calculations.

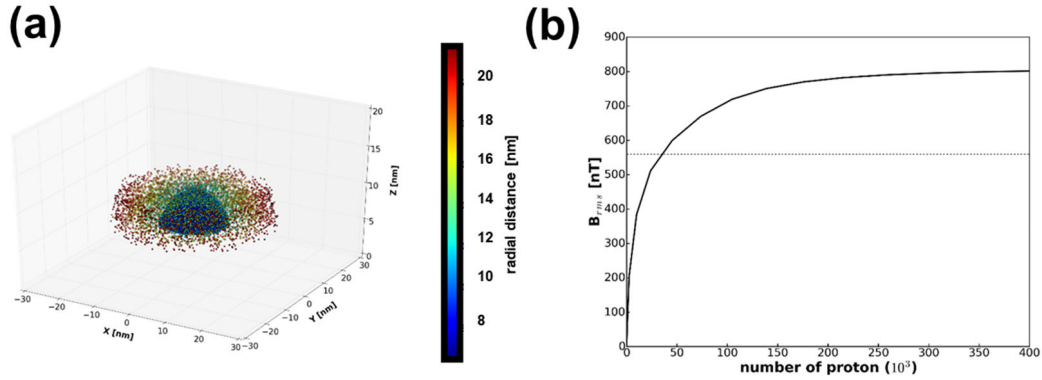

Figure S1(a) 3D mapping of radial distribution of detected nuclear spin from Lipid Bilayer. Different color represents difference in radial distance from NV center. (b) Monte Carlo simulation results of detected $B_{rms}$ signal as function of total detected number of protons. Dotted line shows 70% line from saturation value of $B_{rms}$.

**Effect of standard deviation from fitting of NV center depth on diffusion constant:**

Diffusion constant of $1.5 \pm 0.25$ nm$^2$/μsec at 26.5 °C and $3.0 \pm 0.5$ nm$^2$/μsec at 36.0 °C shown in the manuscript is calculated with NV center depth of 6.6 nm estimated from optimal fitting and an error bar is introduced from uncertainties (e.g. exact thickness of proton layer and exact location) caused by proton layer on the bottom of lipid bilayer.

Additional uncertainty for value of diffusion constant is given by the standard deviation on fitting result for depth of NV center which is $6.6 \pm 0.5$ nm. When NV center depth is 6.1nm, diffusion constant is 0.75 nm$^2$/μsec at 26.5 °C and 1.5 nm$^2$/μsec at 36.0 °C. Whereas when NV center depth is 7.1nm, diffusion constant is 2.5 nm$^2$/μsec at 26.5 °C and 4.75 nm$^2$/μsec at 36.0 °C. Therefore, uncertainty given for diffusion constant shown in the manuscript is at most 1.0 nm$^2$/μsec for 26.5 °C and 1.75 nm$^2$/μsec for 36.0 °C corresponding to ~60% of diffusion constant obtained with the optimal value for NV center depth. Possible improvements on uncertainty of diffusion constant could be achieved by using NV center with depth below 5nm for improved sensitivity of target molecules.

**Calculation of proton density in lipid bilayer and conversion to lipid packing density:**

In DPPC, ~ 80% of protons are contained within hydrocarbon chain of lipids and therefore most of the signal observed in nanoscale NMR is from hydrocarbon chain of lipids. Simple calculation on hydrocarbon chain with volume of 0.913nm$^3$ with 62 proton gives uniform proton density of 68 proton [nm$^{-3}$]. In our manuscript, density of proton in lipid bilayer has been estimated by applying fitting function to Radially Distributed Function (RDF) of MD simulation. Example is shown in Figure S2 below where Integrated RDF of proton in lipid bilayer was fit to a function defined by uniform density σ in equation S3 below. As shown in the figure, the number of protons which exists within radius r from a proton corresponds well with the equation defined by volume with radius r and uniform density σ. S was calculated to be 2.71, close to ~3 which is expected for spherical volume and σ value corresponding to density of proton inside lipid bilayer was 65 proton [nm$^{-3}$]. This result proves density obtained from our method is properly treated as uniform density of proton in lipid bilayer.

$$Total\ number\ of\ proton\ from\ RDF = \sigma \left(\frac{4}{3}\right)\pi r^S \qquad [S3]$$

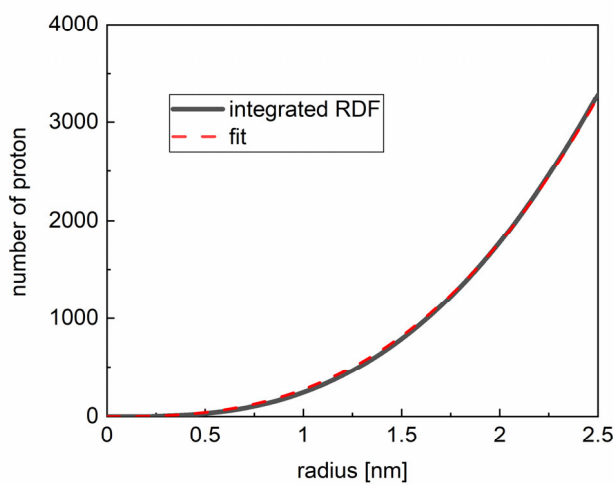

Figure S2 Fitting performed on integrated RDF to estimate density of proton inside lipid bilayer.

Calculated proton density was converted to lipid packing density from equation S4 below. DPPC molecules ($C_{40}H_{80}NO_8P$) contains 80 Hydrogen and lipid bilayers are formed with two phospholipids that are facing each other on fatty chain as shown in Figure 1(b) of main manuscript. Therefore, each lipid containing spot occupies 160 protons. This value was divided by density obtained from nanoscale NMR contrast 65 proton [nm$^{-3}$] multiplied by 5nm thickness assumed for lipid bilayer. This calculation gives area occupied per lipid of 0.49 [nm$^2$/lipid] which is reasonable agreement with 0.47 [nm$^2$/lipid] for DPPC reported in reference 24 of manuscript.

$$\frac{160 \text{ proton}/\text{lipid on both sides}}{65 \text{ proton}/nm^3 \times 5 \text{ nm}} = 0.49 \text{ [nm}^2\text{/lipid]} \qquad [S4]$$

**Effect of using diamond substrate for observation of diffusion constant:**

FRAP measurement result obtained on diamond sample using Rhodamine-B mixed DPPC lipid bilayer was compared with FRAP measurement obtained on mica sample. Both measurements showed recovery of florescence in 5min. which indicates comparable diffusion mechanism between two different substrates. Further study on effect of diamond substrate for lipid bilayer diffusion will be investigated in our future work.

**Details of MD simulation:**

The MD simulations were conducted using the Fujitsu PRIMEGY CX600M1/CX1640M1 (Oakforest-PACS) and SGI Rackable C1102-GP8 (Reedbush). The lipid bilayer systems were constructed using CHARMM-GUI [S1], where the bilayers consist of 200 lipids in a periodic cubic box with a side length of 10 nm perpendicular to the membrane (Figure S3 (a)). As the forcefield parameters, the CHARMM36 [S2] and TIP3P [S3] models were employed for lipid and water molecules, respectively. After energy minimization, the systems were equilibrated for 4.0 ns by applying a Berendsen Barostat [S4] of 1.0 bar and Berendsen thermostat at target temperatures, 298 K, 310 K, and 314 K, respectively. The MD production runs were carried out under canonical conditions using a Nosé – Hoover thermostat [S5] of respective target temperatures with a time constant of 0.5 ps. The electrostatic interactions were evaluated using the particle-mesh Ewald method with a switching function for electrostatic interactions. The electrostatic interactions in real space and Van der Waals interactions were damped to zero with a switching function in the range of 8.5 and 9.0 Å. The intra/intermolecular radial distribution functions were evaluated for all hydrogen atoms in the fatty acid chain, where the hydrogen atoms in the head group and water were excluded.

Integrated RDF was used to calculate the intramolecular and intermolecular closest average proton–proton distances. Typical results for integrated RDF is shown in Figure S3 (b), where the closest proton-to-proton distance was estimated to be 1.88 Å. This value is comparable to the proton distance between two protons bonded to the same carbon on the acidic chain. To calculate the intramolecular closest average proton to proton distance, protons on the acidic chain of a single molecule were used. This is demonstrated in integrated radial distribution functions for a long radius. As demonstrated in Figure S3 (c), when the radius from a single proton approaches a 3 nm, number of protons saturates to a value equal to that of an acid chain part of single DPPC molecule. A similar calculation technique was used to calculate the intermolecular closest average proton to proton distance. In this case, proton from the same DPPC molecule was eliminated in the calculation, and only the contribution from other DPPC molecules was calculated in the RDF.

For rotational analysis, we selected four carbon atoms C21, C216, C31, and C316, where C21 and C31 were neighboring to the headgroup while C216 and C316 were located at the tails of the fatty acid chains. Then, the second rank auto-correlation functions of rotations of two vectors C21-C216 and C31-C316 of DPPC are evaluated using trajectories of 400 ns at 298.15 K and 314.15 K, respectively. The obtained auto-correlation functions were well fitted by a linear combination of three exponential curves, which resulted in larger correlation coefficients than 0.98. All simulations were conducted using the GROMACS version 2018-3. [S6]

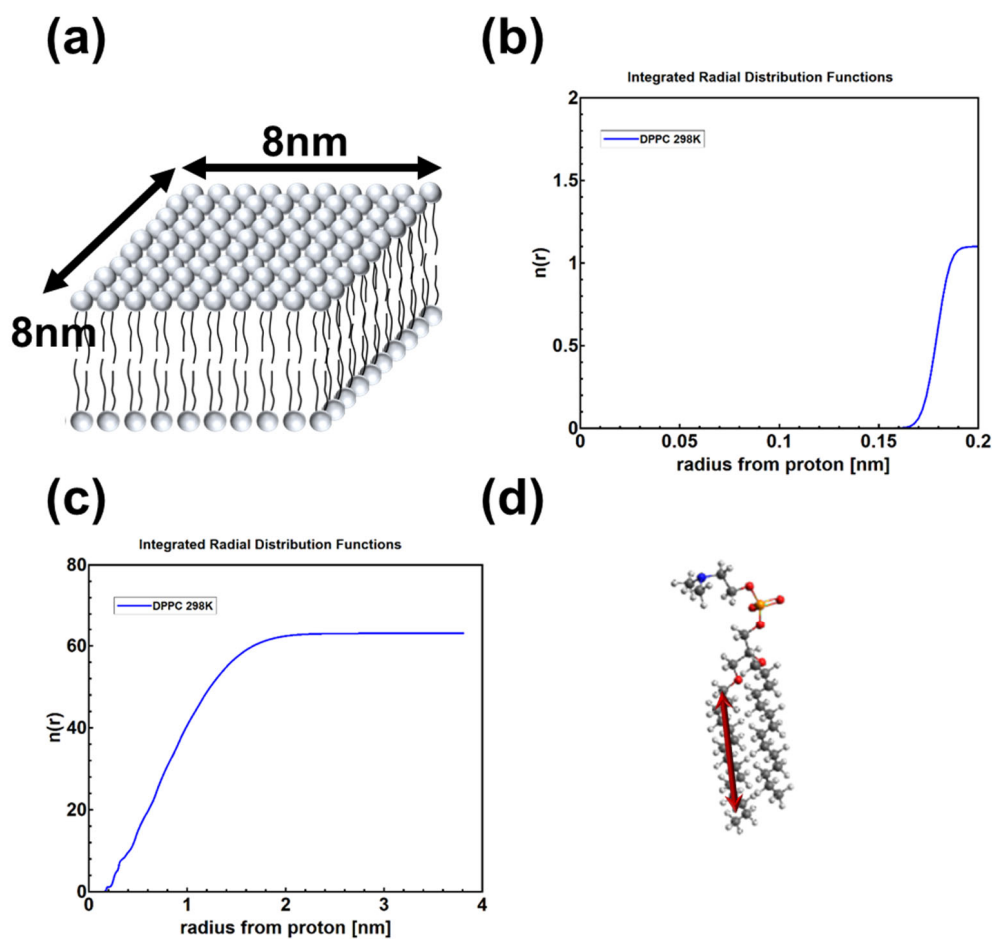

Figure S3: (a) Schematic representation for lipid model used for DPPC MD simulation. 10 lipid x 10 lipid configuration with double layer configuration was used for simulation with total of 200 molecules. (b) RDF calculation for intramolecular closest proton average distance. Calculated by integrated radial distribution functions. Showing 1.88 Å for closest proton. (c) Integrated RDF for larger radius from proton. Showing total number of protons for each lipid. (d) Model used for calculation of rotation of DPPC molecules. Vector connecting first carbon on acidic chain and last carbon (16[th] carbon) on acidic chain is used to analyze rotation and wobble of DPPC molecules.

**2D diffusion model:**

Figure S4 (a) shows a schematic representation of the modeling of 2D diffusion dynamics in the lipid bilayer. As shown in Figure S4 (b), a random starting position was selected within the detection area, and the diffusion area was calculated using eq. S5 to calculate diffusion radius for the change in diffusion time. $D_{lb}$ is diffusion constant of lipid bilayer and $t_{diff}$ is diffusion time. The cross section between the diffusion area and detection area was calculated and divided by the diffusion area to calculate the probability of detection at the corresponding diffusion time. For example, the case shown in Figure S4 (c) represents the case where the detection probability is given as 1 because of the complete overlap between the diffusion area and detection area. Figure S4 (d) shows the case where the probability of detection is given by the overlap between the diffusion area and detection area divided by the diffusion area. The overlap between the diffusion area and detection area was calculated using eq. S6. d is distance between center of detection area and diffusion area and $r_{DV}$ is the radius of detection area. These two cases were applied for all randomly selected starting points, and random points were picked for the number of nuclear spins inside the detection volume.

$$r_{diff} = \sqrt{4D_{lb}t_{diff}} \qquad [S5]$$

$$A(d, r_{diff}, r_{DV}) = r_{diff}^2 \cos^{-1}\left(\frac{d^2 + r_{diff}^2 - r_{DV}}{2dr_{diff}}\right) + r_{DV}^2 \cos^{-1}\left(\frac{d^2 + r_{DV}^2 - r_{diff}}{2dr_{DV}}\right) - \frac{1}{2}\sqrt{(-d + r_{diff} + r_{DV})(d + r_{diff} - r_{DV})(d - r_{diff} + r_{DV})(d + r_{diff} + r_{DV})} \qquad [S6]$$

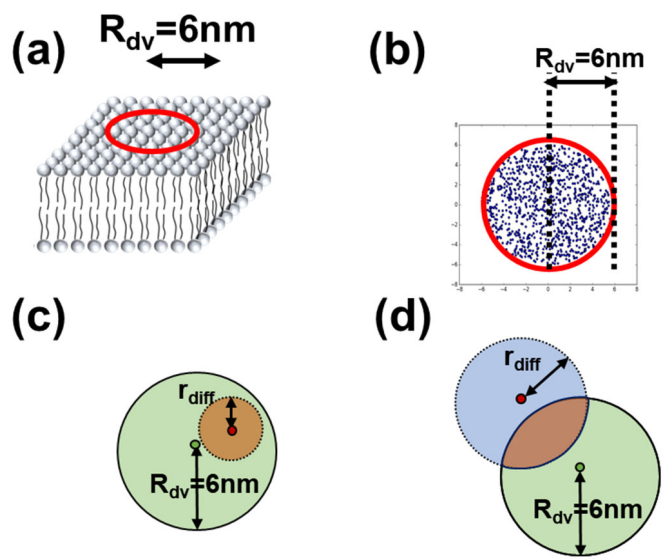

Figure S4: (a) Schematic representation for definition of diffusion area (b) Distribution example of randomly distributed initial point for calculation of diffusion area. (c) Example of calculation of diffusion area and detection area. In the case represented in this figure diffusion area is completely included in detection area. In this case probability of detection is calculated as 1. (d) Example of calculation for cross section of diffusion area and detection area. In this case, overlap between diffusion area and detection area will be divided by area of diffusion area to calculate probability for detection of nuclear spin.